\documentclass[%
 reprint,
nofootinbib,
 amsmath,amssymb,
 aps,
 normalem
]{revtex4-2}

\usepackage{graphicx}
\usepackage{dcolumn}
\usepackage{bm}
\usepackage{ulem}

\usepackage[linktocpage=true]{hyperref}

\renewcommand{\eqref}[1]{Eq. (\ref{#1})}

\usepackage{xcolor}

\newcommand{\mpl}{m_{\mathrm{Pl}}}

\newcommand{\gag}{g_{a\gamma}}
\newcommand{\gagcrit}{g_{a\gamma}^\mathrm{crit}}

\begin{document}

{\hfill KCL-PH-TH/2023-15 \vspace{-20pt}}

\title{Electromagnetic instability of compact axion stars} 

\author{Liina M. Chung-Jukko$^{a}$}
\email{liina.jukko@kcl.ac.uk}
\author{Eugene A. Lim$^{a}$}
\author{David J. E. Marsh$^{a}$}
\author{Josu C. Aurrekoetxea$^{b}$}
\author{Eloy de Jong$^{a}$}
\author{Bo-Xuan Ge$^{a}$}

\vspace{1cm}
\affiliation{${}^a$Theoretical Particle Physics and Cosmology Group, Physics Department, King's College London, Strand, London WC2R 2LS, United Kingdom}
\affiliation{${}^b$Astrophysics, University of Oxford, DWB, Keble Road, Oxford OX1 3RH, United Kingdom}

\begin{abstract}
If the dark matter is composed of axions, then axion stars are expected to be abundant in the Universe. We demonstrate in fully non-linear (3+1) numerical relativity the instability of compact axion stars due to the electromagnetic Chern-Simons term. We show that above the critical coupling constant $\gagcrit\propto M_s^{-1.35}$, compact axion stars of mass $M_s$ are unstable. The instability is caused by parametric resonance between the axion and the electromagnetic field.  The existence of stable compact axion stars requires approximately Planck-suppressed couplings to photons. If the coupling exceeds the critical value, then all stable axion stars are necessarily non-compact. Unstable axion stars decay leaving behind a less massive, less compact, remnant. The emitted radiation peaks at frequency $\omega \sim 1/R_s$, where $R_s$ is the axion star radius.


\end{abstract}

\maketitle



\emph{Introduction:} If dark matter (DM) is composed of axions or axion-like particles (henceforth, axions)~\cite{Marsh:2015xka}, then DM halos are predicted to host an abundance of so-called \emph{axion stars}~(see e.g. Refs.~\cite{Schive:2014dra,Levkov:2018kau,Widdicombe:2018oeo,Eggemeier:2019jsu,Chen:2020cef} for formation mechanism, and Ref.~\cite{Du:2023jxh} for the abundance and merger rates). Axion stars are self-gravitating, time periodic, finite mass solutions of the Klein-Gordon-Einstein equations, which fall under the class of solitonic objects known as oscillatons~\cite{Seidel:1991zh,Seidel:1993zk}. A defining property of axions is that they are real pseudo-scalars, and necessarily couple to gauge fields via the Chern-Simons term. In the case of electromagnetism, this leads to a coupling between the axion and two photons specified by a coupling constant $g_{a\gamma}$ with mass dimension -1. In terms of classical fields, the axion couples to $\vec{E}\cdot\vec{B}$. 

It is known that this coupling can lead to an instability of the axion fields~\cite{Kephart:1986vc,Boskovic:2018lkj,Ikeda:2018nhb}. In particular, within the context of axion stars, this non-linearity is destabilising, as was demonstrated in the weak field perturbative regime in Ref.~\cite{Levkov:2020txo}, and first suggested in Ref.~\cite{Tkachev:1987cd}. In the strong field regime, it was also recently shown that complex scalar boson stars with a coupling to the Chern-Simons term can also become unstable \cite{Sanchis-Gual:2022zsr}.

In this paper, we investigate the stability of compact, relativistic axion stars in the presence of a weak propagating electromagnetic (EM) wave modelling a bath of ambient photons. We use the 3+1 numerical relativity code \textsc{GRChombo} ~\cite{Clough:2015sqa,Andrade:2021rbd,Radia:2021smk}. We find that, as long as (i) the EM wavelength is approximately the size of the axion star and (ii) the coupling exceeds a critical coupling  $\gagcrit \propto M_s^{-1.35}$ where $M_s$ is the axion star mass for fixed axion mass $m$, the star will experience an instability, losing mass via potentially detectable EM emissions.  

To be specific, we find the following:
\begin{itemize}
    \item The instability is induced by {\it parametric resonance}, with an instability band roughly with a bandwidth $\Delta \omega \sim R_s^{-1}$ where $R_s$ is the size of the axion star, centered around $\omega \sim R_s^{-1}$. EM energy is generated exponentially. 
    \item The critical threshold for the coupling is $$\qquad\quad \gagcrit \approx \frac{1.66 \times 10^{-17}}{\mathrm{GeV}} \left[ \left( \frac{M_s}{M_\odot} \right)\left(\frac{m}{10^{-11} \mathrm{eV}} \right)\right]^{-1.35} $$ where we have scaled our results to $m=10^{-11} \mathrm{eV}$ corresponding to  ${\cal O}(M_\odot)$ compact axion stars \cite{Alcubierre:2003sx}.

    \item The timescale of the instability is a power law $$\qquad\quad\tau \propto (\gag- \gagcrit)^{-0.87}$$ and independent of the initial EM seed amplitude.
    \item The instability is largely insensitive to the initial amplitude of the ambient EM field $E_0$ -- since the instability is exponential, the time to trigger it depends on $t_0 \sim \ln E_0$ at best.
\end{itemize}

\begin{figure}[t!]
    \includegraphics[width=\linewidth]{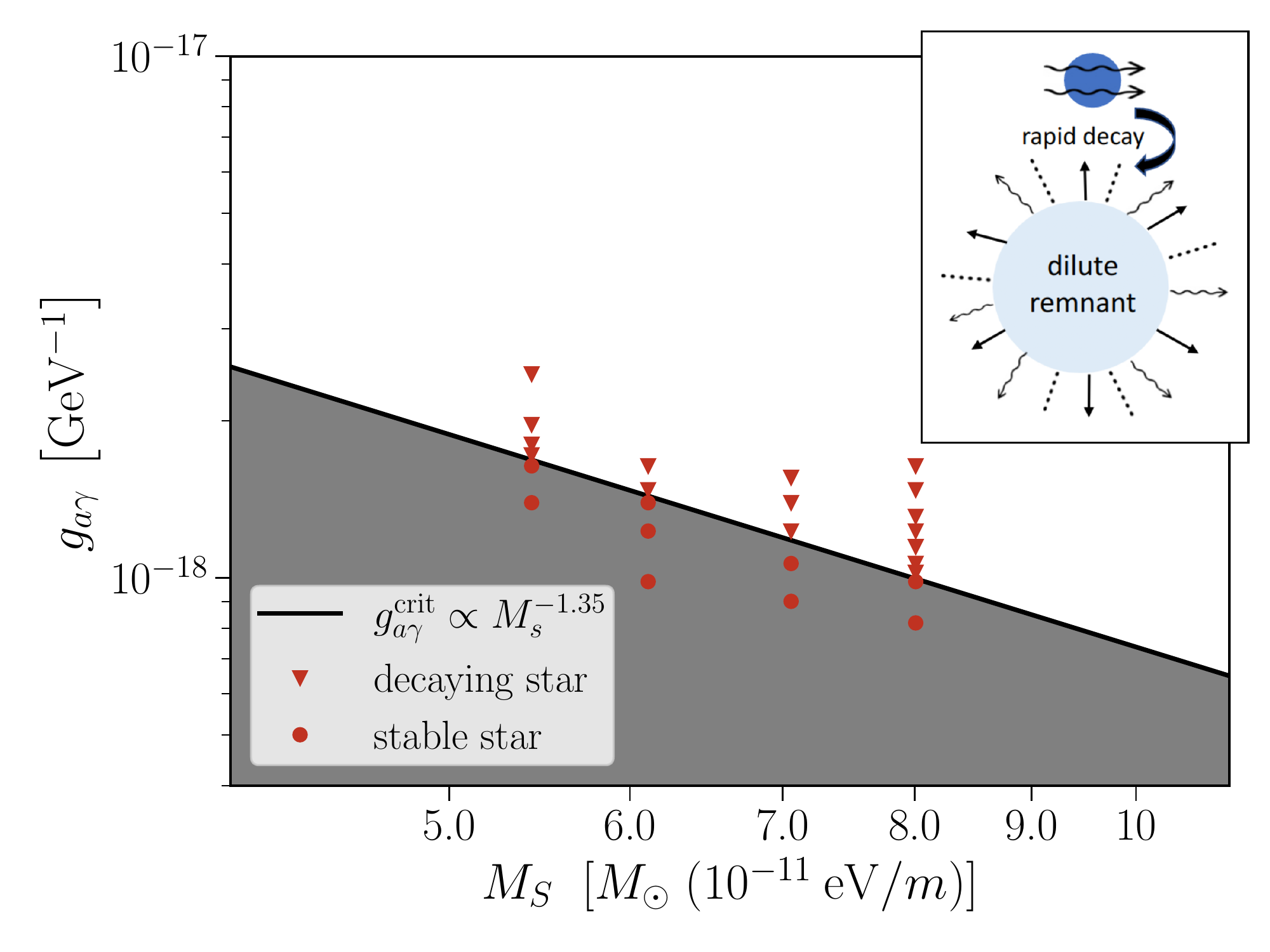}
        \vspace{-0.7cm}
    \caption{The critical coupling (in black) along our simulation data (in red), with triangular simulation points representing a decaying star through scalar, electromagnetic and gravitational radiation (see diagram in top right corner). Our simulations cover $M_s = 0.60, 0.53, 0.46, 0.41 m_{\mathrm{Pl}}^2/m$, and we have plotted the mass ranges scaled to $m=10^{-11}$ eV which correspond to compact axion stars of ${\cal O}(M_\odot)$. }
   
    \label{fig: cartoon}
    \vspace{-0.5cm}
\end{figure}

The presence of this instability forbids axion stars from existing above the critical line $\gagcrit(M_s)$ in the $(M_s,g_{a\gamma})$ plane, as shown in Fig.~\ref{fig: cartoon}.  Compact axion stars have $M_s\sim \mpl^2/m$, and our results imply that stable compact axion stars can exist only if the axion-photon coupling is approximately Planck suppressed (a similar conclusion applies to the axion quartic self-coupling as was shown in Ref.~\cite{Helfer:2016ljl}). 

\emph{Theory:} The electromagnetic field strength tensor and its dual are
\begin{equation} \label{eq: fieldstrength}
F_{\mu \nu}=\partial_{\mu} A_{\nu} - \partial_{\nu} A_{\mu}, \quad \widetilde{F}^{\mu \nu}=\frac{1}{2 \sqrt{-g}} \varepsilon^{\mu \nu \rho \sigma} F_{\rho \sigma},
\end{equation}
with $\varepsilon^{\mu \nu \rho \sigma}$ being the totally antisymmetric Levi-Civita symbol with $\varepsilon^{0123}=+1$. We write the total action\footnote{Our metric signature is $- + + +$, and $\hbar = c =1$.} as 
\begin{equation} \label{eq: action}
\begin{aligned}
S=\int d^{4} x\sqrt{-g}&\left[ \frac{\mpl^2}{16\pi}R-\frac{1}{2} \partial_{\mu} \phi \partial^{\mu} \phi-\frac{1}{2} m^{2} \phi^{2}\right. \\
 &\qquad\left. -\frac{1}{4} F_{\mu \nu} F^{\mu \nu} -\frac{g_{a \gamma}}{4} \phi F_{\mu \nu} \tilde{F}^{\mu \nu} \right],
\end{aligned}
\end{equation}
where $\phi$ is the axion field, and $R$ is the Ricci scalar. The last term in this action is the Chern-Simons term, which acts as a boundary term and hence does not contribute to the stress-tensor. The stress-energy tensor is derived from \eqref{eq: action} to find Einstein's equations $G_{\mu \nu}=8 \pi \mpl^{-2} T_{\mu \nu}$ (see e.g. Ref.~\cite{Gorbar:2021rlt}). 

The equations of motion in the matter sector are 
\begin{align}
\nabla^\mu \nabla_\mu \phi-m^{2} \phi&=\frac{g_{a \gamma}}{4} F_{\mu \nu} \tilde{F}^{\mu \nu}, \label{eq: PhiEOM} \\
\nabla_{\mu} F ^{\mu \nu}&=-g_{a \gamma} J^{\nu},\label{eq: FEOM}
\end{align}
where the current $J^{\nu}$ is defined as $J^{\nu} = \partial_{\mu} \phi \tilde{F}^{\mu \nu}$. The parametric resonance is driven by the EM sector \eqref{eq: FEOM}, as long as the photon frequency is within the resonance band of the axion field. Since the axion oscillates $\omega \sim m \sim R_s^{-1}$, if the photon wavelength is ${\cal O}(R_s)$, resonance will commence.

We solve the full system with numerical relativity using \textsc{GRChombo}~\cite{Andrade:2021rbd, Radia:2021smk, Clough:2015sqa} following the methodology in Ref.~\cite{Zilhao:2015tya,Helfer:2018qgv,Gundlach:2005eh,Palenzuela:2009hx,Hilditch:2013sba}. For a summary, please see appendix \ref{appendix_method}. We construct initial conditions for compact axion stars with ADM masses $0.41 m_{\mathrm{Pl}}^2/m\leq  M_s \leq 0.60 m_{\mathrm{Pl}}^2/m$, which corresponds to 

\begin{equation}
5.4 M_\odot \left(\frac{10^{-11}\mathrm{eV}}{m}\right)\leq  M_s \leq 8.1 M_\odot \left(\frac{10^{-11}\mathrm{eV}}{m}\right)~,\label{eq: massrange}
\end{equation}
following the method used in Ref.~\cite{Helfer:2016ljl,Helfer:2018vtq, Alcubierre:2003sx, Michel:2018nzt,Seidel:1991zh,  Urena-Lopez:2002ptf, Urena-Lopez:2001zjo}. These masses are near the Kaup~\cite{kggeon} limit for black hole formation.

For the EM field initial conditions, we approximate the initial spacetime as Minkowski, since we are interested in the case where the EM field is subdominant to the energy density of the axion star.
This approximation decouples the oscillaton and EM initial conditions from each other, with minimal violations to the initial constraint equations. We choose the components of our gauge field, $A_{\mu} = C_{\mu} e^{i(-k_{\mu}z+\omega_{\mu} t)}$, to describe a single plane wave polarised in the x-direction, with wavevector $k_{\mu}^{(x)} = (\omega^{(x)}, 0, 0, -k^{(x)})$ such that $\omega^{(x)} = k^{(x)}$ initially. We identify $A_0$ and $A_z$ with the gauge mode, and set $C_0 = C_z = C_y = 0$ at the initial time. This ansatz satisfies both the Lorenz gauge $k_{\mu} A^{\mu} = 0$, and the Bianchi identities, which set the dispersion relation for each wave mode.
Using these simplifications, the only non-zero components of the electric and magnetic fields are 
\begin{align}
E_{x} &= \partial_{t} A_{x}-\partial_{x} A_{t}= -\omega^{(x)} C_{x} \sin{(-k^{(x)}z+\omega^{(x)} t)} \\
B_{y} &= \partial_{z} A_{x}-\partial_{x} A_{z} = -\omega^{(x)} C_{x} \sin{(-k^{(x)}z+\omega^{(x)} t)},
\end{align}
where we have used the real part of the gauge fields. We use $k^{(x)} \equiv 2\pi/\lambda \sim 0.10m$, and the amplitude $C_x = 0.001m_{\mathrm{Pl}}$ as our initial conditions. Numerically solving the full non-linear equations to evolve our system implies that all classical backreactions are included in our simulations. Periodic boundary conditions were used throughout the simulations. We show that the constraint equations are satisfied and tested their convergence during evolution in Appendix \ref{sec:convergencetesting}.

\begin{figure*}[t!]
    \centering
    \includegraphics[width=\linewidth]{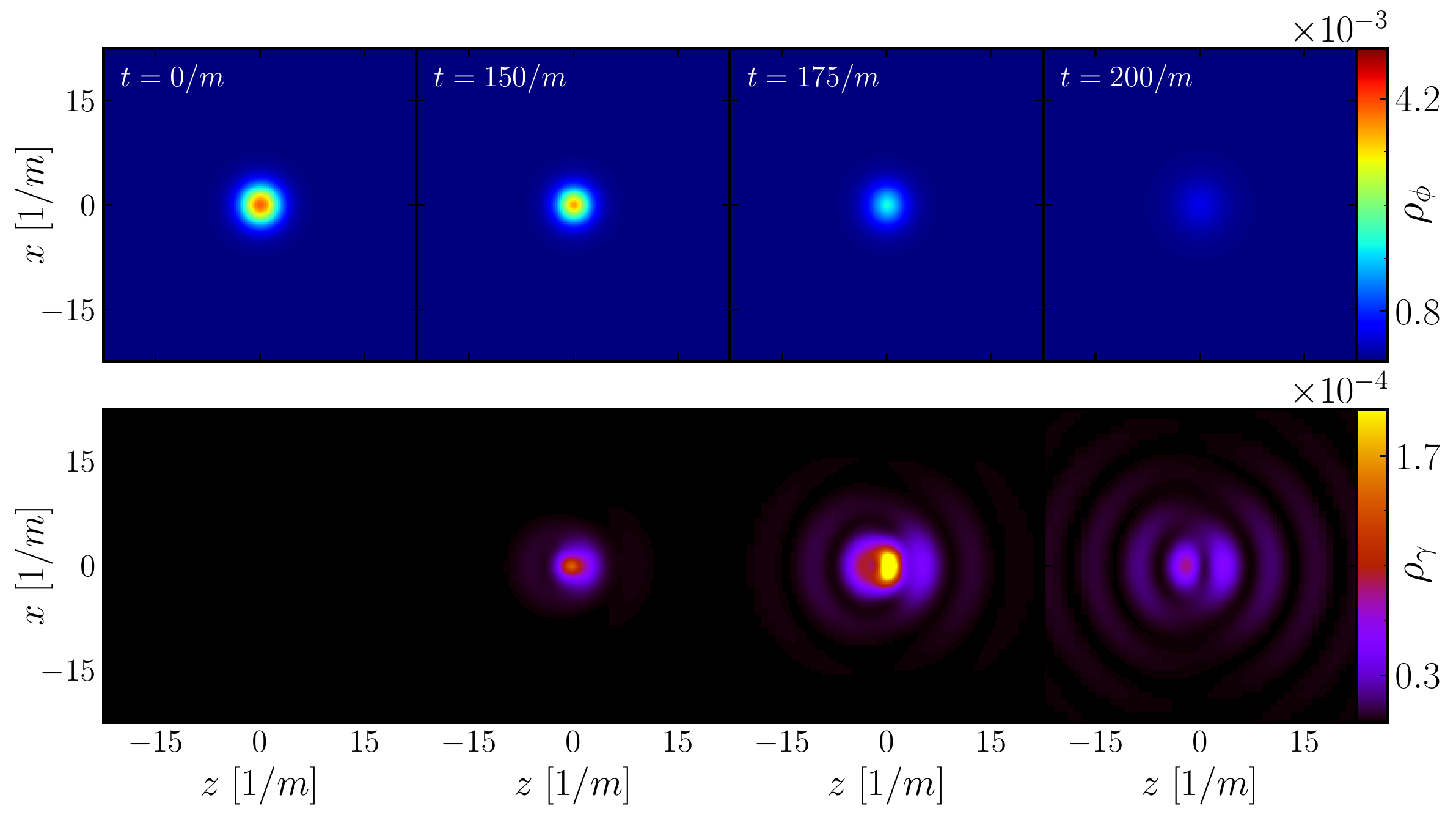}
        \vspace{-0.5cm}
    \caption{Energy densities of the electromagnetic and scalar fields as a slice through the centre of the star for the $ M_s = 0.60 m_{\mathrm{Pl}}^2/m$, $g_{a\gamma}=16m_{\mathrm{Pl}}^{-1}$ case. The EM field (bottom panel), initially polarized in the $x$ direction, is initially propagating from the right to the left. As parametric resonance kicks in, the axion star undergoes rapid dilution and mass loss, with a corresponding burst in the EM energy which is roughly isotropic (see $t=175/m$). The process stops when the axion star dilutes and expands to a size away from the characteristic frequency of the EM spectra. A movie of our simulations for coupling $g_{a\gamma} = 16 \mpl^{-1}$ can be found in \href{https://youtu.be/5UC_PymwAIU}{this link}.}
\label{fig:sliceplots}
    \vspace{-0.5cm}
\end{figure*}


\emph{Results:} 
Slices through our simulation box for the $M_s = 0.60  m_{\mathrm{Pl}}^2/m$, and $g_{a\gamma}=16 m_{\mathrm{Pl}}^{-1}$ case illustrating the evolution of the axion and EM energy density are shown in Fig.~\ref{fig:sliceplots}. An incoming seed EM wave (not visible on the scale shown) causes the axion star to emit a strong burst of EM radiation at $t\sim 100 m^{-1}$. At a later time, $t\sim 200 m^{-1}$, the EM radiation becomes less intense, and the axion star begins to settle into a stable lower mass, less compact, and larger configuration. As the star dilutes and increases in radius $R_s$, its characteristic frequency drifts out of the instability band, shutting down the parametric resonance process.

\begin{figure}
    \centering
    \includegraphics[trim=0.8cm 1.0cm 0.8cm 0.9cm, clip, width=9.0cm]{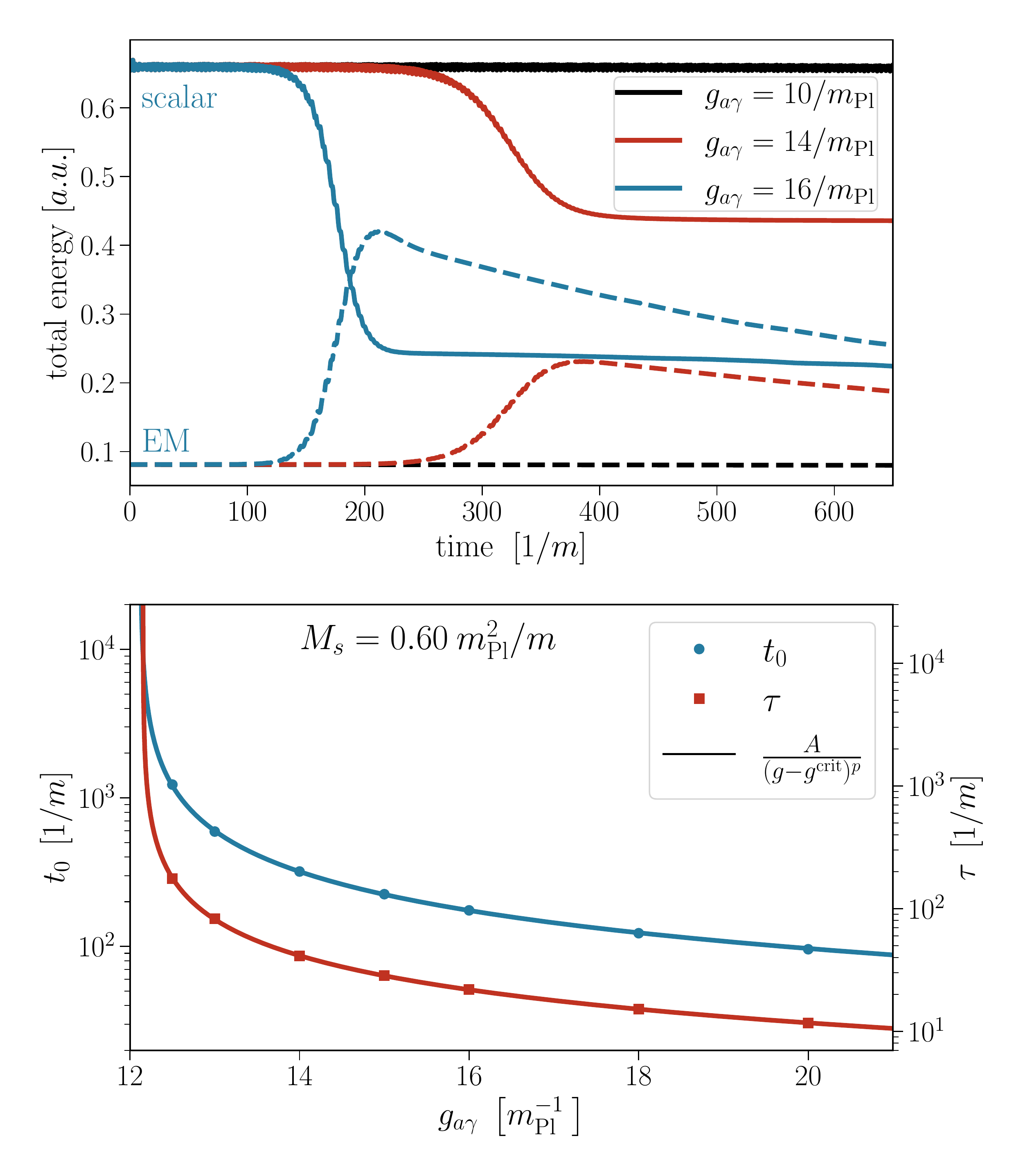}
\caption{\emph{Top:} The total energy in the scalar (solid line) and electromagnetic fields (dashed line) for several values of the coupling $g_{a\gamma}$ for the $ M_s = 0.60 m_{\mathrm{Pl}}^2/m$ case. Total energy conservation (including gravitational energy) is checked by ensuring the Hamiltonian constraint is not violated (see appendix \ref{appendix_method}). \emph{Bottom:} The values of the parameters $t_0$ and $\tau$ from the hyperbolic tangent fit \eqref{eq: tanh} to the EM energy profile as a function of the axion-photon coupling $g_{a\gamma}$. The power law function fitted can be seen in the legend. This gives a critical value for the coupling of $\sim 12.1m_{\mathrm{Pl}}^{-1}$. The simulation errors found through higher resolution runs were of order $0.1\%$ for $t_0$ and $1\%$ for $\tau$. The values for $g^{\mathrm{crit}}$ and $p$ were $12.1m_\mathrm{Pl}^{-1}$ and $0.83$ respectively, for $t_0$, and $12.2m_\mathrm{Pl}^{-1}$ and $0.87$ for $\tau$. }
    \label{fig: t0_and_tau}
        \vspace{-0.5cm}
\end{figure}

We next show in Fig.~\ref{fig: t0_and_tau} (top panel) the time evolution of the total energy in axions and EM radiation, which can be obtained by integrating their respective energy densities (see Appendix \ref{appendix_method}). In order to describe the decay process, we fit a $\tanh$ function for the amplification of the energy of the EM field $E_{\gamma}$:

\begin{equation} \label{eq: tanh}
E_{\gamma}(t) = A \left( \frac{e^{2 (t-t_0)/\tau} + 1}{e^{2 (t-t_0)/\tau} - 1} \right) + B,     
\end{equation}
where the constants $A$ and $B$ depend on the simulation box size.

The amplification in the EM energy sets two timescales: the parameter $t_0$ determines how fast the amplification process begins after the start of the simulation, and $\tau$ can be seen as a measure of the lifetime of the star in the decay process. The dependence of $t_0$ and $\tau$ on the axion-photon coupling $g_{a\gamma}$ is demonstrated in Fig.~\ref{fig: t0_and_tau} (bottom panel); they follow a decaying power law, which has an asymptote at a critical value of $g_{a\gamma} \approx 12.1 m_{\mathrm{Pl}}^{-1}$ based on our simulation data.  We find $\tau \propto g_{a\gamma}^{-0.87}$. We compare this result to the parametric resonance instability timescale for a homogeneous cosmological axion field, which is proportional to $g_{a\gamma}^{-1}$, although the instability is blocked by the expansion of the Universe ~\cite{Preskill:1982cy}. A gravitational potential well, provided by the axion star itself, is required to allow for the instability to develop~\cite{Tkachev:1987cd}. Our results indicate that the decay scaling for relativistic highly  inhomogeneous compact axion stars is comparable but different from the homogeneous case.

We verified the dependence of $t_0$ and $\tau$ on the EM seed amplitude. We find that $\tau$ is independent of the amplitude for fixed photon-axion coupling $g_{a\gamma}$. We note that $A$ from \eqref{eq: tanh} is also constant, indicating that the EM field is amplified by the same amount of energy and hence the amplification has the same shape independent of the amplitude of the EM seed $E_0$. Furthermore, we confirm that $t_0$ has a logarithmic dependence on the initial amplitude of the EM seed, $t_0 \sim \ln{E_0}$ \footnote{As an example, $t_0 \approx -31.3m^{-1} \ln(1860(m_{\mathrm{Pl}}m)^{-1} E_0) + 174m^{-1}$ for $M_s = 0.60 m_{\mathrm{Pl}}^2/m$ and $g_{a\gamma}=15m_\mathrm{Pl}^{-1}$. The constants depend on the value of $g_{a\gamma}$ chosen.}, as the growth of the EM field is an exponential process $E(t) \sim E_0\exp( (t-t_0)/\tau)$. 

Weak field calculations of non-relativistic (and hence non-compact) axion star decay suggest a critical value for the axion-photon coupling $\gagcrit = 7.66 \mpl / \sqrt{8\pi} m M_{s}$, or $\gagcrit \propto M_{s}^{-1}$ \cite{Levkov:2020txo}. While the power law is different, the proximity of the coefficient suggests that decay dynamics are broadly similar in both the weak and strong gravity limits\footnote{This is in agreement with complex scalar boson stars instability found in \cite{Sanchis-Gual:2022zsr}.}. A possible explanation is that it is driven by {\it matter} couplings, with gravity playing only a second order role\footnote{See for example \cite{Amin:2020vja,Amin:2023imi}.}. We also compared the critical mass given in Ref. \cite{Levkov:2020txo} to the remnant axion star mass from our simulations and find they were of the same order of magnitude with our remnants having slightly lower mass. 

\begin{figure}
    \centering
    \includegraphics[width=9cm]{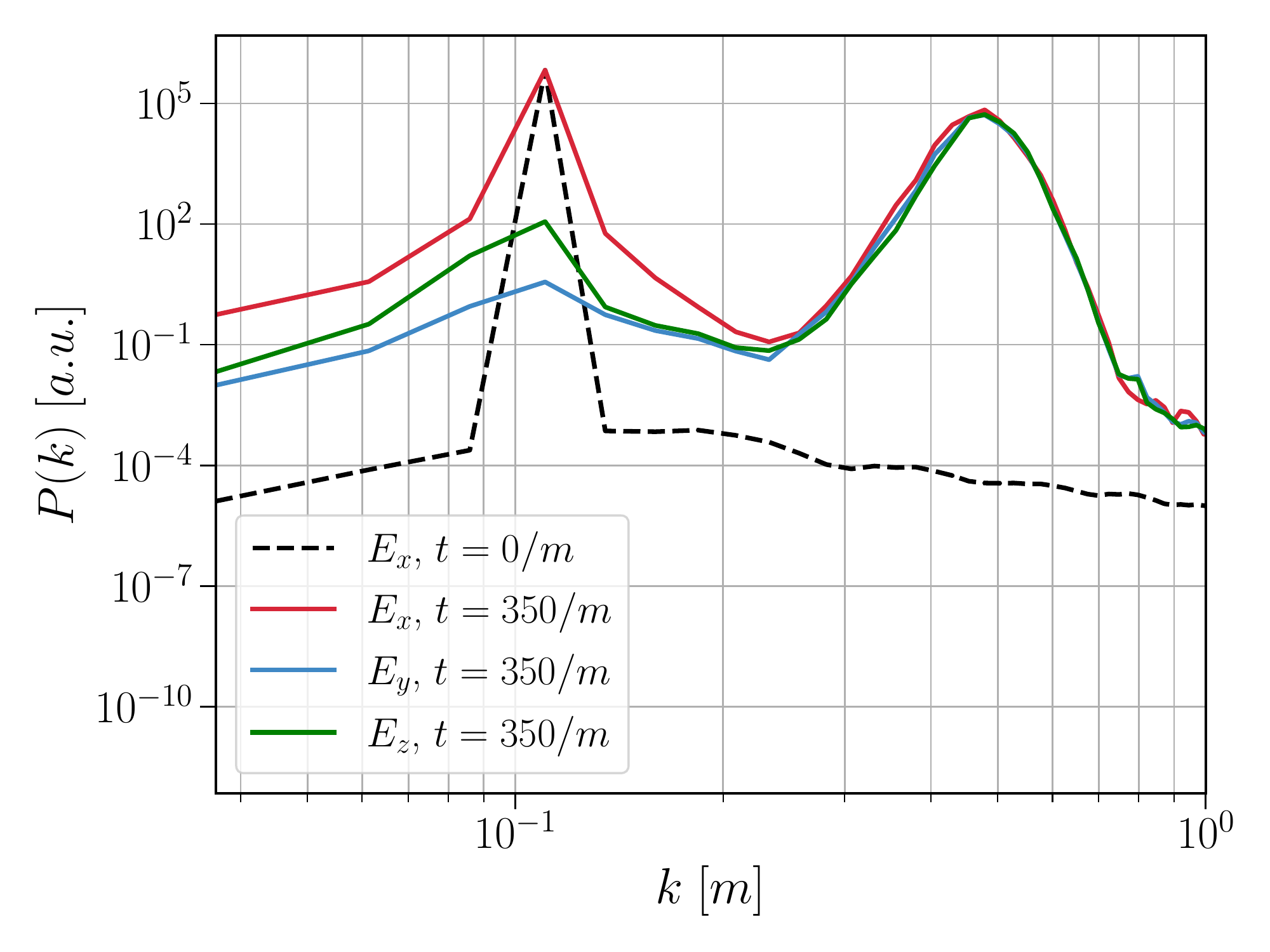}
        \vspace{-0.7cm}
    \caption{Power spectra of the electric field for $g_{a\gamma}=16m_{\mathrm{Pl}}^{-1}$ at time $t=350m^{-1}$, which is after the decay process has ended, for the $ M_s = 0.60 m_{\mathrm{Pl}}^2/m$. The excitation of the wave mode corresponding to the inherent frequency scale of the axion star around $k \sim 0.5 m$ is clearly visible, where the energy is equipartitioned. The black dashed line demonstrates the power spectrum of the initial EM seed, polarised in the $x$ direction.}
    \label{fig: powerspectra}
        \vspace{-0.5cm}
\end{figure}

In Figure \ref{fig: powerspectra}, we show the power spectra $P(k)$ of the $x$, $y$ and $z$ components of the electric field at $t=350m^{-1}$ for $g_{a\gamma}=16m_{\mathrm{Pl}}^{-1}$, after the decay process has happened, along with the original seed (black dashed line), demonstrating the frequency of the EM radiation emitted by the axion star as it decays. We obtained the power spectrum by performing a fast Fourier transform on the spatial electric field, and then integrating the square of the transform in $k$-space. We note two salient points. First, around the incoming frequency $k=0.1m$, the $E_x$ power broadens, with a corresponding smaller power in $E_z$ and $E_y$ power, but no large amplification. Secondly, the primary power of the emission lies around $k\sim 0.5m$, equipartitioned between the $x$, $y$ and $z$ components. This scale corresponds to the diameter $2R_s$ of the axion star $2kR_s\sim 2\pi, k \sim 0.6m$, capturing the emission from parametric resonance. This equipartition of energies arises from (i) the total momentum of the system must remain small as the initial EM waves carry negligible momenta (ii) the source current for the radiation  is the initially spherically symmetric axion star $J_{\mu} \propto \partial_\mu \phi$.  We note that while we saw evidence of birefringence between the $x$ and $y$ components before decay, post-decay this effect is subdominant. We leave to future work a complete study of the emission power and possibly circularly polarised emission due to $CP$ violation.

\emph{Conclusions:} We have demonstrated in fully non-linear simulations that axion stars are unstable above a critical line $\gagcrit \propto M_s^{-1.35}$ in the plane of mass and coupling constant, exploding into EM radiation. Crucially,  we have shown this decay process with a log dependence on the amplitude of the plane wave, suggesting that ambient radiation alone would be sufficient to destabilise compact axion stars on Hubble timescales. As an example, assuming the resonant band $\delta \omega \sim m$ and $\gag = 15m_\mathrm{Pl}^{-1}$, destabilisation of ${\cal O}{(M_\odot})$ compact axion stars stimulated by the Cosmic Microwave Background photons will take approximately $t_0 \sim 0.05$ seconds, smaller than the Hubble expansion time by many orders of magnitude. 

Populations of axion stars can form cosmologically due to mergers of dark matter halos or from collapse of cosmological perturbations~\cite{Widdicombe:2018oeo}, with a computable rate~\cite{Du:2023jxh}. Our results suggests that if $g_{a\gamma}>\gagcrit$, compact axion stars quickly decay into EM radiation which can heat the intergalactic medium, with potentially observable consequences~\cite{Escudero_in_prep}. Conversely, this will impact attempts to search for these objects via gravitational waves from their mergers \cite{Chia:2020psj,Toubiana:2020lzd}. In future work, we intend to study the multi-messenger gravitational, scalar, and EM radiation of axion star decays.

\emph{Acknowledgements:} We would like to thank Thomas Helfer for his early contribution to the project. We also thank members of the \textsc{GRChombo} Collaboration for technical support and help. LMCJ is supported by a studentship funded by the Science and Technologies Facilities Council (UK). DJEM is supported by an Ernest Rutherford Fellowship from the Science and Technologies Facilities Council (UK). JCA acknowledges funding from the Beecroft Trust and The Queen’s College via an extraordinary Junior Research Fellowship (eJRF). This work used the DiRAC@Durham Cosma facility managed by the Institute for Computational Cosmology on behalf of the STFC DiRAC HPC Facility (www.dirac.ac.uk), under DiRAC grant ACTP238. The equipment was funded by BEIS capital funding via STFC capital grants ST/P002293/1, ST/R002371/1 and ST/S002502/1, Durham University and STFC operations grant ST/R000832/1.

\bibliography{bosonstar}

\clearpage

\appendix

\section{ Numerical methodology} \label{appendix_method}

Our numerical implementation on \textsc{GRChombo} \cite{Andrade:2021rbd, Radia:2021smk, Clough:2015sqa} evolves the gravity sector using the CCZ4 formalism \cite{Alic:2011gg, Alic:2013xsa}, together with the integrated moving puncture gauge \cite{Campanelli:2005dd,Baker:2005vv}. The decomposition of the matter sector is based on \cite{Helfer:2018qgv} (see Appendix A in that paper in particular), with an additional Chern-Simons coupling between the EM and scalar sector $$L_\mathrm{CS} = -\frac{g_{a \gamma}}{4}\phi F_{\mu \nu} \tilde{F}^{\mu \nu} $$ and no gauge coupling (i.e. $e=0$).\\

We use as diagnostic quantities the energy densities in both the scalar $\rho_\phi=n^\mu n^\nu T^{\phi}_{\mu\nu}$ and the EM field $\rho_\gamma=n^\mu n^\nu T^\gamma_{\mu\nu}$, which are obtained projecting their respective energy momentum tensor with the normal vector $n^\mu$ to the three-dimensional hypersurface. In terms of the fields, these are expressed as
\begin{align}
T_{\mu\nu}^{\phi} &= \nabla_{\mu}\phi\nabla_\nu\phi - \frac{g_{\mu\nu}}{2}\left(\nabla_\sigma\phi\nabla^{\sigma}\phi + m^2\phi^2\right)~, \\
T_{\mu\nu}^{\gamma} &= F_{\mu\alpha}F^{\alpha}{}_{\nu} -\frac{1}{4}g_{\mu\nu}F_{\alpha\beta}F^{\alpha\beta}~.
\end{align}
Note that the Chern-Simons term does not contribute to the stress-tensor as it is topological. The total energy in axions and the EM field can then be calculated integrating the energy densities over a volume $V$
\begin{equation}
    E_{\{\phi,\gamma\}} =\int_{V} \sqrt{-g} \rho_{\{\phi,\gamma\}} dV~.
\end{equation}

\section{Convergence testing}  \label{sec:convergencetesting}

We monitor the evolution of the average Hamiltonian and momentum constraints in a sphere of radius $64m^{-1}$ centered around the axion star. For the initial conditions of the (subdominant) EM field, we approximate the initial spacetime as Minkowski, which introduces minimal violations to the constraints that are quickly damped via CCZ4. In addition, we checked that the gauge field constraint violation was negligible and under control throughout the simulations.

\begin{figure}[t!]
\centering
\includegraphics[width=\linewidth]{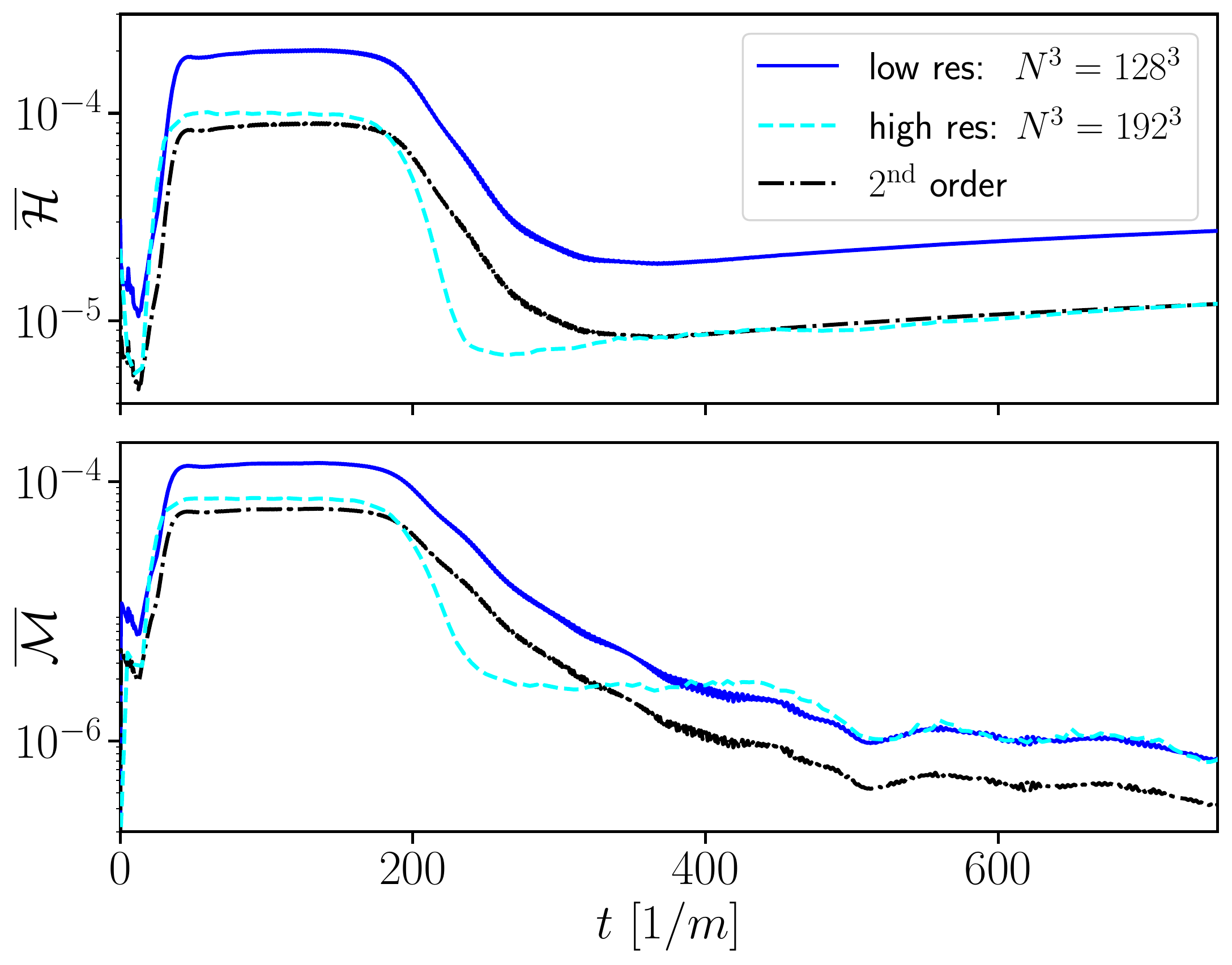}
\caption{Convergence test of the average Hamiltonian $\overline{\mathcal{H}}$ and momentum constraint $\overline{\mathcal{M}}$ violations in a sphere of radius $64 m^{-1}$ centered around the axion star. Simulation box size is $L=256m^{-1}$, and low and high resolution runs have $N^3=128^3$ and $N^3=192^3$ number of coarse grid points, respectively.}
\label{fig:convergence}
\end{figure}

We tested convergence for the $\gag = 16m_{\mathrm{Pl}}^{-1}$ case comparing the evolution of the Hamiltonian and momentum constraint violations for two resolutions in a simulation with box size $L=256m^{-1}$ and 7 refinement levels. For the low and high resolutions, we increased the number of coarse grid points from $N^3 = 128^3$ to $N^3=192^3$, resulting in finest grid sizes of $\Delta x\approx 1.6\times 10^{-2} m^{-1} $ and $\Delta x\approx 10^{-2} m^{-1}$, respectively\footnote{Note that the size of the axion star is $R_s\sim m^{-1}$, implying that it is resolved with $\mathcal{O}(100)$ number of grid points.}. In Fig. \ref{fig:convergence} we show how the errors in the Hamiltonian and momentum constraints were reduced by a factor consistent with 2nd order convergence when the resolution was increased. 

One might worry about the self-amplification of the EM field due to the periodic boundary conditions. By doubling the size of the simulation box, we tested that the evolution of the EM field was not altered by the reflected energy density until the resonance phase had finished (but can spoil the 2nd order convergence of the constraints as shown in Fig \ref{fig:convergence}).

Additionally, we found that the post-decay electromagnetic energy density in Fig. \ref{fig: t0_and_tau} drops due to the fact that the standard resolution used in our simulations cannot track the highest frequency modes generated during the resonant phase. The Kreiss-Oliger dissipation used in \textsc{GRChombo} to remove the noise introduced by regridding also removes EM modes with wavelengths of the order of the grid spacing. We checked that these higher modes can be recovered by increasing the resolution of the simulation, and estimated that the errors in the physical variables characterising the amplification of the EM field (see Fig. \ref{fig: t0_and_tau}) are negligible -- order $1\%$ at most. For the case of non-decaying stars, we evolved the system until $t \approx 1500/m€$.

\clearpage

\end{document}